\documentstyle[epsfig,11pt]{article}

\def\beq{\begin{equation}}
\def\eeq{\end{equation}}

\def\noi{\noindent}
\def\lra{\leftrightarrow}

\title{\bf Loops in the gluon emission amplitude: reggeization from the
Glauber scattering}
\author{M.Braun \\
Department of high-energy physics,\\
University of S. Petersburg, 198904 S. Petersburg, Russia }
 \date{December 2008}

\begin{document}
\maketitle
\medskip
\noi{\bf Abstract}
It is shown that in  the Glauber scattering of a fast quark
in the external field loop corrections to the gluon emission amplitude due
to virtual softer gluon  after
renormalization coincide with a correction
due to reggeization of the exchanged gluon in the BFKL picture.

\section{Introduction}
In the perturbative QCD the inclusive gluon production is one of the most
important observables, which can be directly compared to the experimental data.
Applied to the interaction with heavy nuclei it was first studied in
~\cite{bra1} using the standard AGK rules and neglecting emission from the
triple pomeron vertex. Later Yu. Kovchegov and K.Tuchin derived the single inclusive
cross-section from the dipole picture ~\cite{kovch}. Their expression contained an
extra term as compared to ~\cite{bra1} which was shown to correspond to emission from
the triple-pomeron vertex ~\cite{bra2}. However recently J.Bartels, M.Salvadore and G.-P.
Vacca performed a new derivation based on the dispersion approach, which studies discontinuities
of amplitudes in various $s$-channels ~\cite{BSV}. Their results are different from the previous ones
in ~\cite{kovch}. They contain new terms of a complicated structure, which seem to involve
the so-called BKP states, higher pomerons composed of 3 and 4  reggeized gluons.
Presence of such states  will inevitably make calculations of the inclusive cross-section
much more difficult (if possible at all), since their wave functions  depend
on many variables and are unknown at present. So the problem to understand the origin
of the difference in the two derivations is quite important.

It may be that after all the initial physical picture in the two approaches is
different. In the Bartels picture one is studying discontinuities of the standard Feynman
diagrams in the Regge kinematics. In the Kovchegov-Tuchin picture the gluon emission
occurs due to Glauber scattering of fast quarks and gluons passing through the
nucleus with instantaneous interactions with its components. However one can prove that
tree diagrams which describe single and double gluon emission coincide in these two
pictures.  The inclusive cross-sections also include diagrams with loops formed by the
gluon softer than the observed one. In this paper we
demonstrate that for single gluon emission such loops are also correctly given
by the Kovchegov-Tuchin technique. In the BFKL approach the mentioned loop generates
the Regge trajectory of the exchange gluon and thus is responsible for the gluon
reggeization. Our results show that gluon reggeization is in fact also realized by the
loop contributions to the emission during Glauber scattering.

As a result it looks as if the physical picture employed in the two different approaches
is fully equivalent. Then the difference in the results either is spurious, with
additional terms found in ~\cite{BSV} actually cancelling, or is  a consequence of the
difference in the derivation of the inclusive cross-sections one of which should be
in error. We do not know the answer to this question and leave it to future studies.

\section{The diagrams}
\begin{figure}
\hspace*{6 cm}
\epsfig{file=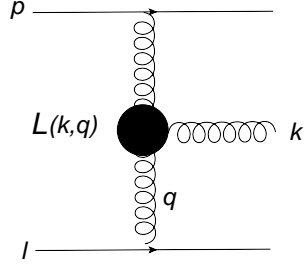,width=4 cm}
\caption{Emission from the Lipatov vertex.}
\label{fig1}
\end{figure}

\begin{figure}
\hspace*{1 cm}
\epsfig{file=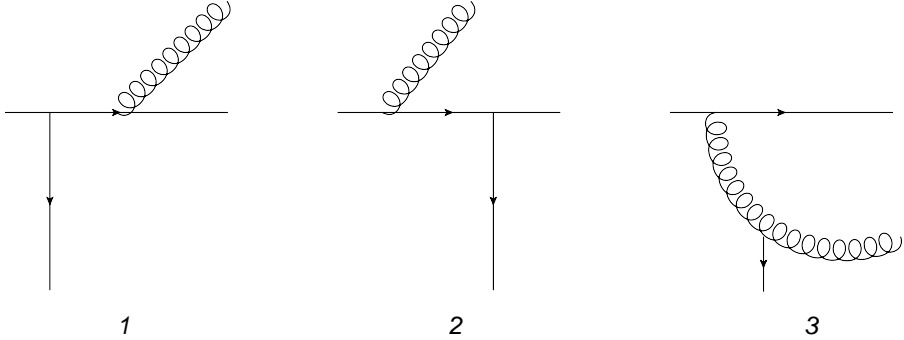,width=12 cm}
\caption{Lowest order diagrams for gluon emission in the external field.}
\label{fig2}
\end{figure}

We shall study emission of a gluon of momentum $k$ in a collision of two quarks with momenta
$p$ and $l$. Quark masses will be taken equal to zero, so that $p_-=p_\perp=l_+=l_\perp=0$.
We take the axial gauge with the gluon field $G$ satisfying $Gl=0$. In the BFKL approach, in
the lowest order, the amplitude is described by the "Lipatov vertex"
\beq
L_{cba}(k,q)=2\Big(\frac{(ke)}{k_\perp^2}-\frac{(k+q,e)}{(k+q)_\perp^2}
\Big)f^{cba},
\label{lip}
\eeq
(see Fig. \ref{fig1}).
In the next order the exchanged gluon reggeizes and
the Born amplitude $A^{(0)}$ is multiplied by the sum of two gluon Regge trajectories
\beq
\omega(k+q)(Y-y)+\omega(q)y,
\eeq
where
\beq
\omega(q)=Ng^2\int\frac{d^2\kappa_\perp}{16\pi^2}
\frac{q_\perp^2}
{\kappa_\perp^2(q-\kappa)_\perp^2}
\label{om}
\eeq
and $Y$ and $y$ are the rapidities of the projectile quark and
emitted gluon respectively. One can interpret both $Y-y$ and $y$
as the result of integration over the longitudinal momentum of the
virtually emitted gluon:
\beq
Y-y=\int^{p_+}_{k_+}\frac{d\kappa_+}{\kappa_+},\ \
y=\int^{k_+}\frac{d\kappa_+}{\kappa_+}.
\eeq
From these expressions it follows that $\omega(k+q)(Y-y)$ is formed
by virtual emissions of gluons harder that the observed one and
$\omega(q)y$ is formed by emissions of gluons softer that the observed one.
We shall be interested in the latter case, so that the loop correction
will be given just by
\beq
A^{(0)}\omega(q)y,
\label{loopcor}
\eeq
where
\beq
\omega(q)y=\int\frac{d\kappa_+}{2\kappa_+}\int \frac{d^2\kappa_\perp}{8\pi^3}\
Ng^2\frac{q_\perp^2}{\kappa_\perp^2(q-\kappa)_\perp^2}.
\label{omy}
\eeq
Our aim is to compare this expression with the one which is obtained in the
Kovchegov-Tuchin picture, when one introduces virtual corrections to the
three lowest order diagrams shown describing emission during Glauber rescattering
and shown in Fig. \ref{fig2}.
It is assumed here that both the fast quark and gluon interact with the target at
a certain light-cone "time" $x_+\equiv t=0$ with an instantaneous interaction,
corresponding to an  exchange of a gluon with a purely transverse momentum $q=q_\perp$.
In the lowest order it is trivial to show that the sum of the three diagrams shown in
Fig. \ref{fig2} gives the same Born amplitude $A^{(0)}$ in Fig. \ref{fig1}. The problem is to
study the next order and compare it to (\ref{loopcor}).

\begin{figure}
\hspace*{1 cm}
\epsfig{file=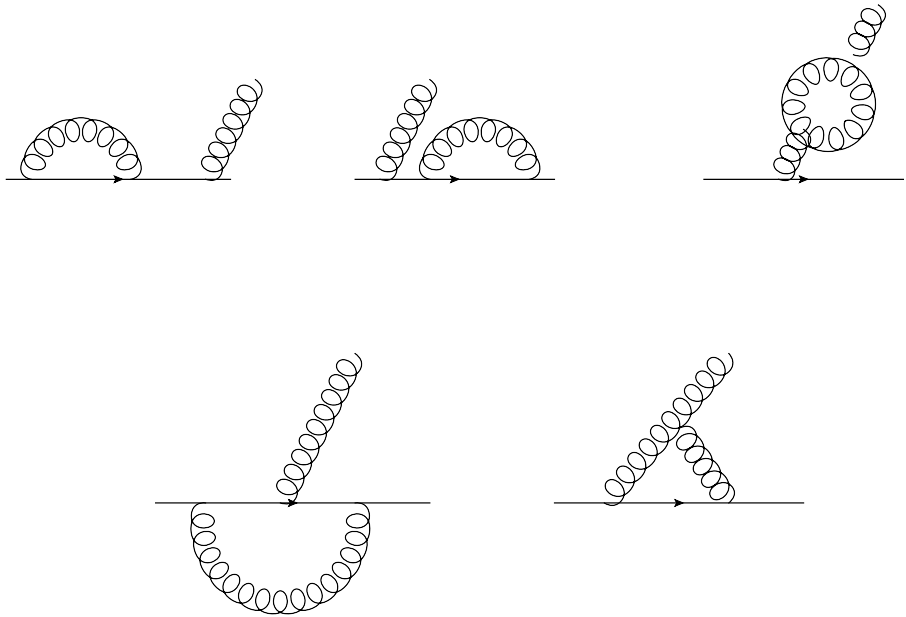,width=12 cm}
\caption{Diagrams with gluon emission and self-masses and vertex parts.}
\label{fig3}
\end{figure}

\begin{figure}
\hspace*{5 cm}
\epsfig{file=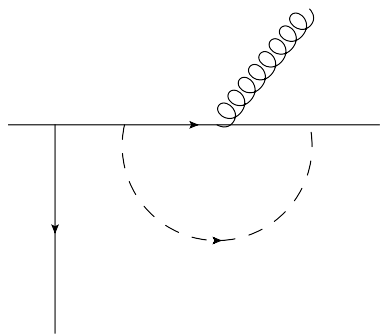,width=5 cm}
\caption{Diagram with gluon emission from the quark, which vanishes.}
\label{fig4}
\end{figure}

All diagrams with such corrections can be obviously obtained from
the five diagrams for gluon emission without interaction with either
the self-mass insertions or nontrivial vertex function shown in Fig. \ref{fig3}.
The desired diagrams will follow if we allow any of the 6 particles
involved to once interact with the target. This generates 30 diagrams.
However, as we shall see, 9 of them with mass insertions into the external lines
are to be dropped, which leaves 21 diagrams. Further reduction of their
number will follow from the requirement of the logarithmic character of
the longitudinal integration and from the color factors in limit
$N_c\to\infty$. In particular we find in this limit
\beq
t^ct^at^c=0,
\eeq
which eliminates diagrams like shown in Fig. \ref{fig4} from the start.

\begin{figure}
\hspace*{1 cm}
\epsfig{file=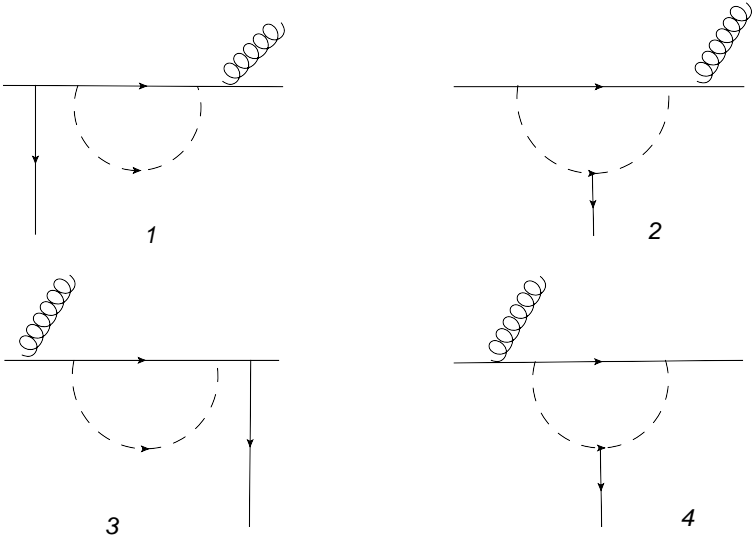,width=10 cm}
\caption{Diagrams (A1)-(A4) for gluon emission  proceeding from the quark self-mass.}
\label{fig5}
\end{figure}

\begin{figure}
\hspace*{1 cm}
\epsfig{file=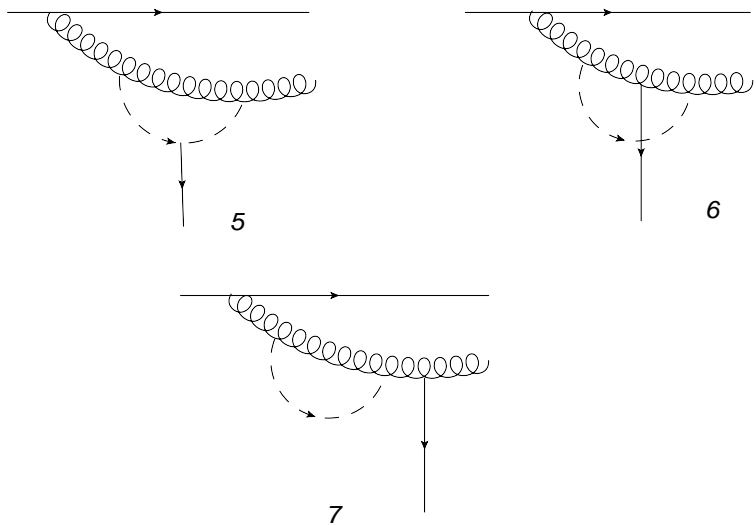,width=10 cm}
\caption{Diagrams (A5)-(A7) for gluon emission  proceeding from the gluon self-mass.}
\label{fig6}
\end{figure}

\begin{figure}
\hspace*{1 cm}
\epsfig{file=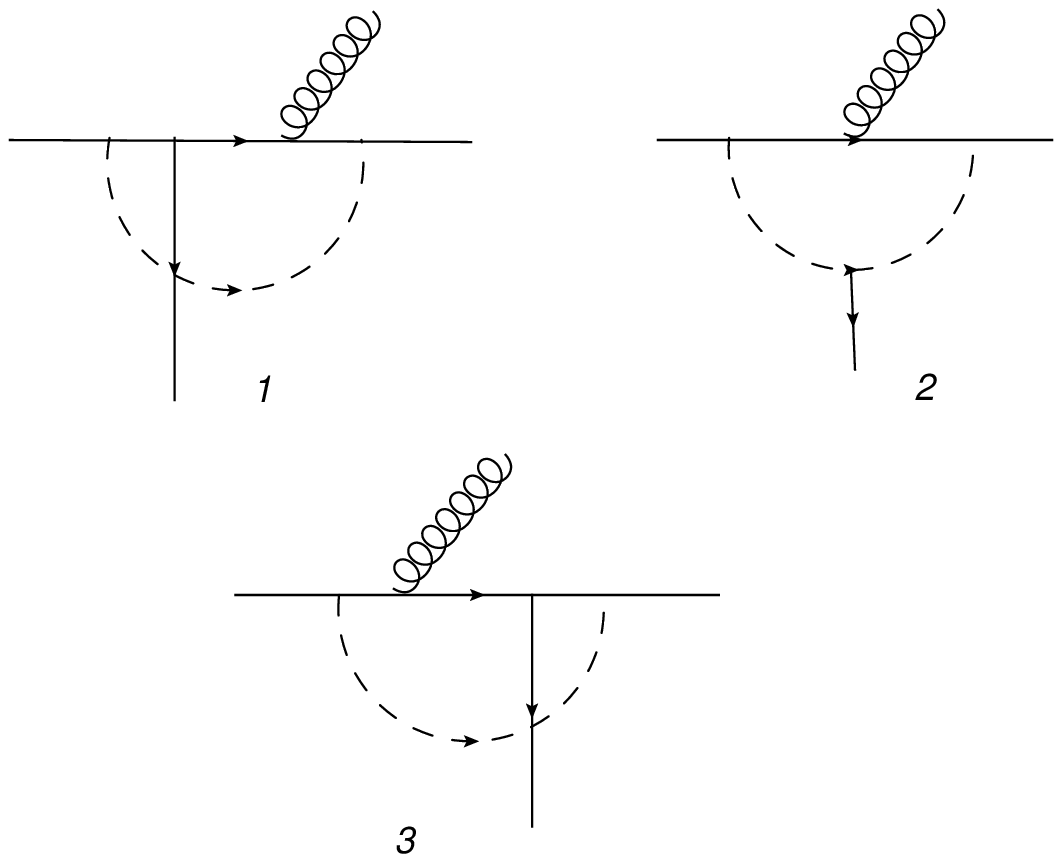,width=10 cm}
\caption{Diagrams (B1)-(B3) for gluon emission  proceeding from the vertex part.}
\label{fig7}
\end{figure}

\begin{figure}
\hspace*{1 cm}
\epsfig{file=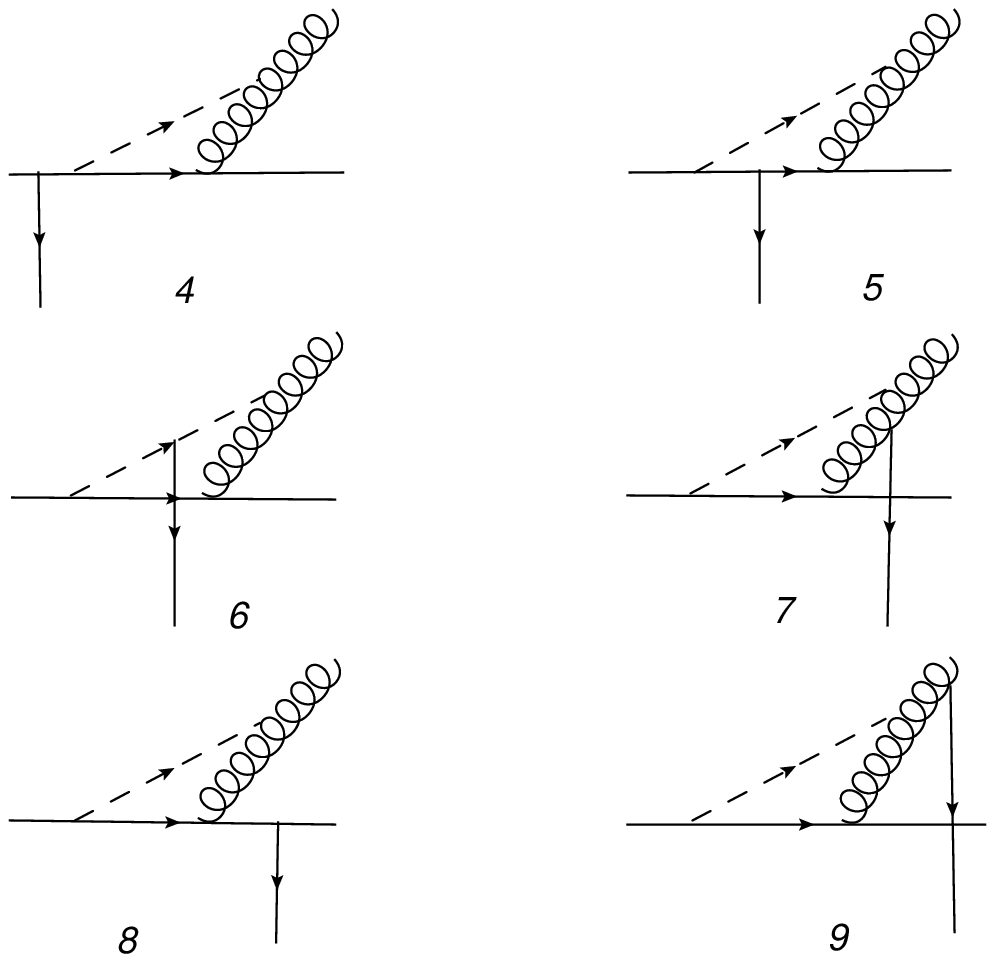,width=12 cm}
\caption{Diagrams (B4)-(B9) for gluon emission  proceeding from the vertex part.}
\label{fig8}
\end{figure}

The remaining diagrams are shown in Figs. \ref{fig5}-\ref{fig8} . Figs. \ref{fig5} and \ref{fig6}
collect diagrams (A1)-(A7) proceeding from the quark and gluon self masses.
Figs. \ref{fig7},\ref{fig8} show diagrams (B1)-(B9) which come from the two vertex parts.
In these diagrams the harder gluon is shown with a wiggly line and the softer one
with a dashed line. To  the vertex diagrams (B4) -(B9)
one should add similar ones with hard and soft gluon interchanged, which we denote as
(B41),(B51) etc, and also diagrams in which in these latter diagrams the direction of the soft
gluon momentum is reversed, denoted as (B42), (B52) etc.

\section{Technical instruments}
\subsection{Basic elements}
We recall that we use the formalism in which interactions with the
target occur at given  "times" $t$, which are in fact the
light-cone coordinate $x_+$.  A convenient
technique is provided by the old time-ordered perturbation
theory in which intermediate particles are on shell and their
propagators are integrated out to give standard energy factors,
which we denote as $T$-factors.

We consider a case when the transition includes a single
instantaneous interaction $V$ acting at $t=0$. Then
the full $S$-matrix is
\beq
S(\infty,-\infty)=-iU(\infty,0)VU(0,-\infty).
\eeq
Its matrix element is given by
\beq
<\alpha|S|\beta>=-i<\alpha|U(\infty,0)|\beta_1><\beta_1|V|\alpha_1>
<\alpha_1|U(0,-\infty)|\beta>,
\eeq
where
\beq
<\alpha|U_n(0,-\infty)|\beta>=(-i)^n\frac{1}{E_\alpha-E_\beta}
<\alpha|iL_I|\gamma_1>\prod_{k=2}^n
\frac{<\gamma_{k-1}|iL_I|\gamma_k>}{E_{\gamma_{k-1}}-E_\beta}
\label{ufinl1}
\eeq
and
\beq
<\alpha|U_n(\infty,0)|\beta>=(-i)^n\frac{1}{E_\beta-E_\alpha}
<\alpha|iL_I|\gamma_1>\prod_{k=2}^n
\frac{<\gamma_{k-1}|iL_I|\gamma_k>}{E_{\gamma_{k-1}}-E_\alpha}.
\label{ufinl2}
\eeq
Here $L_I$ is the interaction Lagrangian for the gluon field,
$|\alpha_1>$, $|\beta_1>$ and $\gamma_k>$ are intermediate states over which summation
is to be done.
Note that in
the relativistic normalization the phase volume contains denominators $2k_+$
for each particle with momentum $k$. Part of this factors cancels by appropriate
relativistic $\delta$-functions, so finally there appears exactly one such
denominator for each particle propagator in the Feynman picture.

The initial projectile and target quarks with momenta $p$ and $l$ respectively
move along the $z$-axis, with $p_-=l_+=0$ and both $p_+$ and $l_-$ are
assumed to be equal and large. We shall study emission of a gluon with
momentum $k$ such that $k_+<<p_+$. The loop correction will proceed from
emission of a softer gluon with momentum $\kappa$ such that $\kappa<<k_+$.
One is to integrate over $\kappa$
\beq
\int\frac{d\kappa_+d^2\kappa_\perp}{(2\pi)^32\kappa_+},
\eeq
which operation will not be indicated explicitly in the following.
For the gluons we use the axial gauge with the numerator in the propagator
\beq
h_{\mu\nu}(k)=g_{\mu\nu}-\frac{k_\mu l_\nu+l_\mu k_\nu}{(kl)}.
\eeq
Interaction will be carried by an exchange of a purely transverse
gluon with momentum $q=q_{\perp}$ and the propagator numerator
$l_\mu p_\alpha/(pl)$,
where $\mu$ refers to the projectile and $\alpha$ to the target.
In accordance with the Glauber picture it is also assumed that the
"+"-component of the travelling particle is not changed by the interaction.
Due to this form of the interaction the numerator in the gluon propagator
with arbitrary number of interactions with the target has a simple form
\beq
H_{\mu\nu}(k_1,k_2,...k_n)=H_{\mu\nu}(k_1,k_n)=
g_{\mu\nu}-\frac{k_{n\mu} l_\nu+l_\mu k_{1\nu}}{(kl)}+
\frac{l_\mu l_\nu (k_1k_n)}{(kl)^2}.
\label {heq}
\eeq
Here $k_1,k_2,...k_n$ are the successive gluon momenta during its
interactions and it is assumed that $k_{1+}=k_{2+}=...=k_{n+}$.
The emitted gluon has a polarization vector $e(k)$ which satisfies
\beq
(e(k)k)=(e(k)l)=0,\ \ {\rm so\ \ that}\ \ e_+=0,\ \
e_-=-\frac{k_{\perp}^2}{k_+}.
\eeq
If the emitted gluon interacts with the target,
so that successive gluon momenta from its emission until its observation
are $k_1,k_2,...k_n,k$ with $k_{1+}=k_{2+}=...k_+$, then the
polarization vector $e(k)$ is changed to
\beq
E(k_1)=e(k)-l\frac{(k_1e(k))}{(k_1l)}
\label{eedef}
\eeq
with the properties
\beq
(E(k_1)k_1)=(E(k_1)l)=0,\ \ {\rm so\ \ that}\ \ E_+=0,\ \
E_-=-\frac{k_{1\perp}^2}{k_+},\ \ E_\perp=e_\perp.
\eeq
Note that both (\ref{heq}) and (\ref{eedef}) include additional factors
$1/2k_+$ which appear as the external field gives rise to new gluon propagators.
So in the end exactly one such factor appears for the propagating gluon whether
in the external field or not.

In the projectile quark each matrix $\gamma_\mu$ which corresponds to
some interaction is changed to $2p_\mu$. In the target quark a similar
matrix $\gamma_\alpha$ is changed to $2l_\alpha$.

From these rules one concludes that if the soft gluon is attached to
the target quark, its propagation is described by the factor
\beq
Q(\kappa_1,\kappa_2)=
p^\mu p^\nu H_{\mu\nu}(\kappa_1,\kappa_2)=
\frac{p_+^2}{\kappa_+^2}(\kappa_1\kappa_2)_{\perp},
\label{qfac}
\eeq
where $\kappa_1$ and $\kappa_2$ are the gluon momenta with which it interacts
with the quark and it is assumed that $\kappa_{1+}=\kappa_{2+}=\kappa_{+}$.

\begin{figure}
\hspace*{5 cm}
\epsfig{file=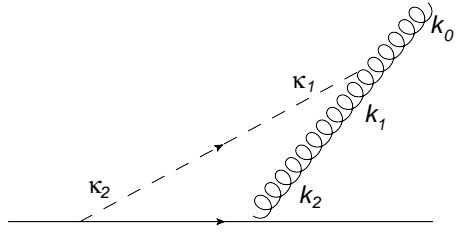,width=6 cm}
\caption{Gluon emission  from the 3-gluon vertex.}
\label{fig9}
\end{figure}

As a part of our diagrams the emission vertex with interaction between
hard and
soft gluon enters shown in Fig. \ref{fig9}, where also Lorentz indexes of the
participating gluons are indicated. Coupled to the projectile quark
and emitted gluon polarization vertex it is given by a factor
\beq
V=p_{\mu}H^{\mu\nu}(\kappa_2,\kappa_1)\Gamma_{\nu\sigma\rho}
(\kappa_1,k_1,-k_0)
H^{\xi\sigma}(k_2,k_1)p_{\xi}E^\rho(k_0),
\eeq
where $k_0=k_1+\kappa_1$

Taking into account that $\kappa_+<<k_+$
and retaining only the leading and the next-to-leading terms we find
(see a sketch of the derivation in Appendix 1)
\[
V=-2\frac{p_+^2}{\kappa_+^2}(\kappa_1\kappa_2)_\perp(k_2e)_{\perp}
\]\beq
+2\frac{p_+^2}{\kappa_+k_+}\Big((k_1\kappa_2)_{\perp}(k_2e)_\perp-
(\kappa_1k_2)_{\perp}(\kappa_2e)_{\perp}+
(\kappa_2k_2)_{\perp}(\kappa_1e)_{\perp}\Big).
\label{v1}
\eeq
Interchange of the gluons $k\leftrightarrow \kappa$ leads to a change of sign
\beq
V_1=V(k\leftrightarrow \kappa)=-V
\label{v}
\eeq

\begin{figure}
\hspace*{5 cm}
\epsfig{file=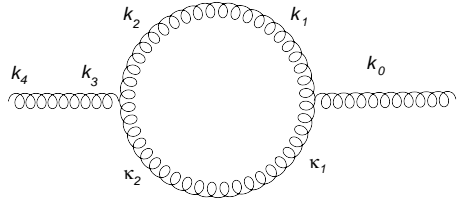,width=6 cm}
\caption{Gluon emission  from the gluon self-mass.}
\label{fig10}
\end{figure}

The final element we have to know is the momentum part of
the diagram with emission of the observed gluon from the
gluon polarization operator, formed by the hard and soft gluons
shown in Fig. \ref{fig10}, with $k_{i+}=k_+>>\kappa_{1+}=\kappa_{2+}=\kappa_+$,
$i=0,1,2,3,4$. It is given by
\beq
\Pi=p^\alpha H_{\alpha\beta}(k_4,k_3)\Gamma_{\beta\xi\mu}(k_3,k_2,\kappa_2)
H_{\mu\nu}(\kappa_2\kappa_1)H_{\xi\sigma}(k_2,k_1)
\Gamma_{\nu\sigma\rho}(\kappa_1,k_1,k_0)E^\rho(k_0).
\label{pidef}
\eeq

A sketch of its calculation is given in Appendix 2. The leading and next-to-leading
terms in $1/\kappa_+$ of $\Pi$ are found to be
\[
\Pi=4\frac{p_+k_+}{\kappa_+^2}(\kappa_1\kappa_2)_{\perp}(k_4e)_{\perp}
-4\frac{p_+}{\kappa_+}(k_4e)_{\perp}\Big((k_1\kappa_2)_{\perp}+
(k_2\kappa_1)_{\perp}\Big)
\]\beq
+8\frac{p_+}{\kappa_+}\Big(
(\kappa_1k_4)_{\perp}(\kappa_2e)_{\perp}-
(\kappa_2k_4)_{\perp}(\kappa_1e)_{\perp}\Big).
\label{pi}
\eeq

\section{Calculation of the  diagrams}

In this section we find the final expressions for the contribution
of our diagrams. As mentioned, in writing them out we suppress  the sign of integration
over the soft gluon momentum
\[ \int\frac{d\kappa_+}{2\kappa_+}\int\frac{d^2\kappa_{\perp}}{8\pi^3}\]
and also factor $q_{\perp}^2$ supplied by the interaction and  factors
$g$. Interaction with the target quark contributes a common factor
\[it^b\frac{2(pl)}{p_+}\]
of which we retain only the momentum part $2(pl)/p_+$.

Generally the contributions of each diagram splits into three factors:
the momentum factor $M$, the colour factor $C$ and the energetic factor $T$
coming from the time integrations.   We denote
$a$ and $b$ the colors of the emitted gluon and the one interacting
with the target quark respectively.

For the following the numerical coefficients and signs are of importance.
We take the vertexes as they are given by Feynman rules from $iL_I$
for each gluon emission. According to (\ref{ufinl1}) and (\ref{ufinl2})
this brings in an overall factor $(-i)$
for each such interaction. The interaction with the target $V=iL_0$ consists
of three factors:  two vertexes for the gluon emission and absorption
and a gluon propagator in the momentum space, the latter depending only
on the transverse momentum. Each vertex comes with factor $i$ and the
gluon propagator carries factor $(-i)$. Of these we preserve only the
$i$ for the gluon emission from the projectile, the other two $i$'s
cancelling each other. So the total overall factor is given by $(-i)^n$
where $n$ is the total number of interactions excluding the one with the
target, with all interactions including the one with the target give by $iL$.
For all the diagrams with a loop it gives $(-i)^3=i$. For the tree
diagrams it gives $-i$.

For a particular diagram each interaction with the quark supplies
factor $i$.  Each
gluon line supplies factor (-1).
The interaction with the target involves the 3-gluon vertex with all
lines outgoing:
\beq
-f^{abc}\Gamma_{\mu\nu\rho}(k_1,k_2,k_3),
\label{3gout}
\eeq
where the gluons $(k_1,\mu,a)$
$(k_2,\nu,b)$ and $(k_3,\rho,c)$ are counted anti-clockwise.
So effectively
each 3-gluon vertex
accompanying the interaction with the target supplies (-1), so that
interactions with the target do not produce additional factors $(-1)$.
The 3-gluon vertexes for gluon emission are taken with all lines incoming.
They have the opposite sign with respect to (\ref{3gout}) and so do not
give any new factors $(-1)$.
The two 3-gluon vertexes in the gluon self mass  supply factor (-1),
since one of them includes reflected momenta.

\subsection{Lowest order}

The corresponding diagrams, which we denote as (1),(2) and (3)are shown in Fig. \ref{fig2}.

{\bf 1.} The initial expression for the momentum part is
\[M=i\bar{u}\hat{e}\hat{p}\hat{l}u
\cdot\frac{1}{2p_+l_-}\cdot\frac{2(pl)}{p_+}.\]
The last factor comes from the target.
Calculating the matrix element gives
\[\bar{u}\hat{e}\hat{p}\hat{l}u=4(ep)(pl)=-4(pl)p_+\frac{(ke)_\perp}{k_+},\]
so in the end
\[M=-4i(pl)\frac{(ke)_\perp}{k_+}.\]
The energetic factor is
\[T=-\frac{1}{k_-}=\frac{2k_+}{k_\perp^2}.\]
The colour factor is $t^at^b$,
so we find
\beq
(1)=-8it^at^b(pl)\frac{(ke)_\perp}{k_\perp^2}.
\eeq

{\bf 2.} The $M$ factor is the same, the $T$ factor has the opposite sign and the
in the $C$-factor the order of $t$'s is reversed. So
\beq
(2)=+8it^bt^a(pl)\frac{(ke)_\perp}{k_\perp^2}.
\eeq

{\bf 3.} Now
\[M=-\bar{u}\hat{e}(k_1)u\cdot\frac{2(pl)}{p_+}
=-4(pl)(pe)=4(pl)\frac{(k_1e)_\perp}{k_+}.\]
The $T$-factor is
\[T=\frac{1}{k_{1-}}=-\frac{2k_+}{k_{1\perp}^2}.\]
The colour factor is
\[C=f^{acb}t^c=-f^{abc}t^c.\]
So finally
\beq
(3)=-8(pl)f^{abc}t^c\frac{(k_1e)_\perp}{k_{1\perp}^2}.
\eeq

Summing all three contributions we find the standard Lipatov
expression graphically shown in Fig. 1
\beq
(1)+(2)+(3)=8(pl)f^{abc}t^c\Big(\frac{(ke)_\perp}{k_{\perp}^2}-
\frac{(k_1e)_\perp}{k_{1\perp}^2}\Big).
\label{123}
\eeq

\subsection{Diagrams generated by the self-masses (Figs. \ref{fig5} and \ref{fig6})}

{\bf A1.}
\[
M=-i\bar{u}\hat{e}\hat{p}\gamma^\mu\hat{p}
\gamma^\nu\hat{p}\frac{\hat{l}}{l_-}u
h_{\mu\nu}(\kappa)\frac{1}{(2p_+)^3}\cdot \frac{2(pl)}{p_+}.\]
Calculating the matrix element and
using  (\ref{qfac}) we get
\[ M=-i4(pl)(pe)\frac{1}{p_+^3}\frac{p_+^2}{\kappa_+^2}
\kappa_{\perp}^2=
4(pl)(ke)_{\perp}\frac{\kappa_\perp^2}{\kappa_+^2}.\]
The colour factor is
\[ C=t^at^ct^ct^b=\frac{N}{2}t^at^b .\]
The $T$ factor can be readily read from the diagram:
\beq
T=\frac{1}{k_-^2(\kappa_--k_-)}=\frac{1}{k_-^2\kappa_-}+
\frac{1}{k_-\kappa_-^2}.
\eeq
The first term  has order $\kappa_+$ and does not lead to
a logarithmic contribution.  So for our purpose
\[T=-\frac{8\kappa_+^2k_+}{\kappa_{\perp}^4k_{\perp}^2}.\]
Combining all factors we get
\beq
(A1)=-16Ni(pl)t^at^b\frac{(ke)_{\perp}}{k_\perp^2}\frac{1}{\kappa_\perp^2}
\label{a1}
\eeq

{\bf A2.} Similar calculations give
\[
M=-i^4\bar{u}\hat{e}\hat{p}\gamma^\mu\hat{p}
\gamma^\nu\hat{p}u
H_{\mu\nu}(\kappa_1,\kappa_2)\frac{1}{(2p_+)^2}\cdot \frac{2(pl)}{p_+}
\]\[
-4(pe)\frac{p_+^2}{\kappa_+^2}\frac{1}{p_+^3}
(\kappa_1\kappa_2)_\perp=
4(pl)(ke)_{\perp}\frac{(\kappa_1\kappa_2)_\perp}{\kappa_+^2}.\]
The colour factor is
\[ C=-t^at^dt^cf^{bdc}=
\frac{1}{2}iNt^b.
\]
The time factor, read from the diagram, is
\[ T=8\frac{\kappa_+^2k_+}{\kappa_{1\perp}^2\kappa_{2\perp}^2k_\perp^2}.
\]
Combing all factors we get
\beq
(A2)=16iNt^at^b(pl)\frac{(ke)_\perp}{k_\perp^2}
\frac{(\kappa_1\kappa_2)_\perp}{\kappa_{1\perp}^2\kappa_{2\perp}^2}=
16iNt^at^b(pl)\frac{(ke)_{\perp}}{k_\perp^2}
\frac{[\kappa(\kappa+q)]_\perp}{\kappa_{\perp}^2(\kappa+q)_{\perp}^2}.
\label{a2}
\eeq

{\bf A3.} The $M$-factor is the same as in (A1). The $T$-factor has the opposite sign.
In the colour factor the order of $t^a$ and $t^b$ will be reversed. So
we find
\beq
(A3)=16iN(pl)t^bt^a\frac{(ke)_{\perp}}{k_\perp^2}\frac{1}{\kappa_\perp^2}.
\label{a3}
\eeq

{\bf A4.} The $M$ factor is the same as in (A2). The $T$-factor hs the opposite sign.
In the colour factor the order of $t^a$ and $t^b$ will be reversed. So
we find
\beq
(A4)=-16iNt^bt^a(pl)\frac{(ke)_{\perp}}{k_\perp^2}
\frac{[\kappa(\kappa+q)]_\perp}{\kappa_{\perp}^2(\kappa+q)_{\perp}^2}
\label{a4}
\eeq

{\bf A5.} Now the momentum part is given via the gluon self-mass insertion $\Pi$
shown in Fig \ref{fig10} with $k_4=k_3\to k_2$:
\[
M=-2\Pi\frac{1}{(2k_+)^2}\cdot \frac{2(pl)}{p_+}.
\]
The time-factor given  has order $\kappa_+^2$, so that
it is sufficient to take $\Pi$ from (\ref{pi}) in the leading order
\[
M=-4(pl)\frac{(k_2e)_{\perp}}{k_+}\kappa_+^2(\kappa_1\kappa_2)_{\perp}.
\]
The colour factor is
\[ t^cf^{dcc_1}f^{c_1bc_2}f^{c_2ad}=-\frac{1}{2}Nf^{bac}t^c.\]
The $T$-factor is
\[
T=-8\frac{\kappa_+^2k_+}{\kappa_{1\perp}^2\kappa^{2\perp^2}k_{2\perp}^2}.
\]
Combining all factors we get
\beq
(A5)=-16N(pl)f^{bac}t^c\frac{(k_2e)_{\perp}}{k_{2\perp}^2}
\frac{(\kappa_1\kappa_2)_{\perp}}{\kappa_{1\perp}^2\kappa_{2\perp}^2}
=
-16N(pl)f^{bac}t^c\frac{[(k+q)e]_{\perp}}{(k+q)_\perp^2}
\frac{[\kappa(\kappa+q)]_{\perp}}{\kappa_{\perp}^2(\kappa+q)_\perp^2}.
\label{a5}
\eeq

{\bf A6.}
The momentum part is given via the gluon self-mass insertion $\Pi$
shown in Fig \ref{fig10} with $k_4=k_3$ and $\kappa_1=\kappa_2$:
\[
M=-2\Pi\frac{1}{(2k_+)^2}\cdot \frac{2(pl)}{p_+}=
-4(pl)\frac{(k_3e)_{\perp}}{k_+}\frac{\kappa_{\perp}^2}{\kappa_+^2}.
\]
The colour factor is found to be the same as in (A5).
The $T$-factor is
\[
T=-8\frac{\kappa_+^2k_+}{\kappa_{\perp}^4k_{3\perp}^2}
\]
So we finally get
\beq
(A6)=-16N(pl)f^{bac}t^c\frac{(k_3e)_{\perp}}{k_{3\perp}^2}
\frac{1}{\kappa_{\perp}^2}=
-16N(pl)f^{bac}t^c\frac{[(k+q)e]_{\perp}}{(k+q)_{\perp}^2}
\frac{1}{\kappa_{\perp}^2}.
\label{a6}
\eeq

{\bf A7.}
In this case the $M$ factor is given by the self-mass insertion
$\Pi$ with $k_2\to k_1$, and $k_4=k_3=k_0\to k_2$.
\[ M=-(pl)\Pi\frac{1}{p_+k_+^2}.\]
The $T$-factor has a leading term of the order $\kappa_+$:
\[ T=-8\frac{\kappa_+k_+^2}{\kappa_\perp^2k_{2\perp}^4}+
8\frac{\kappa_+^2 k_+k_{1\perp}^2}{\kappa_\perp^4 k_{2\perp}^4},\]
so that  we have to retain  terms of the order
$1/\kappa_+$ in (\ref{pi}).
We find
\[\Pi T=32p_+k_+^2\frac{(k_2e)_\perp}{\kappa_\perp^2 k_{2\perp}^4}
\Big(k_{1\perp}^2+2(k_1\kappa)_\perp\Big)
=32p_+k_+^2\frac{(k_2e)_\perp}{\kappa_\perp^2 k_{2\perp}^2}
\Big(1-\frac{\kappa_\perp^2}{k_{2\perp}^2}\Big)\]
The colour factor is twice greater than in (A5) and (A6):
\[ C=f^{cc_2c_1}f^{c_1c_2d}f^{dba}t^c=-Nf^{bac}t^c. \]
So in the end
\beq
(A7)=
32N(pl)f^{bac}t^c\frac{[(k+q)e]_{\perp}}{(k+q)_{\perp}^2}
\frac{1}
{\kappa_{\perp}^2}\Big(1-\frac{\kappa_\perp^2}{(k+q)_{\perp}^2}
\Big).
\label{a7}
\eeq

The second term in this expression
\beq
32N(pl)f^{bac}t^c\frac{[(k+q)e]_{\perp}}{(k+q)_{\perp}^4}
\eeq
corresponds to a contribution $\propto 1/(k+q)^4$ in the 4-dimensional
language, since $(k+q)_-=(p-p'_-\simeq 0$. This contribution is eliminated
by the gluon self-mass renormalization which requires it to vanish on the mass shell.

\subsection{Diagrams generated by vertexes (Figs. \ref{fig7} and \ref{fig8})}

Diagrams (B1)-(B3) have the $T$-factor $\propto\kappa_+^3$, so that they do not
lead to the logarithmic integration in $\kappa_+$ and can be neglected.
To the rest of the diagrams shown in Figs. \ref{fig7} and \ref{fig8}
and denoted as (B4), (B5),...,
new ones should be added with soft and hard gluon interchanged,
denoted correspondingly as (B41), (B51),.... Furthermore to these latter
diagrams one should add ones with the direction of the soft gluon momentum
reversed: $\kappa\to -\kappa$. These last diagrams will be denoted as
(B42), (B52),...

{\bf B4.}
The momentum part is
\[M= \bar{u}\gamma^\xi \hat{p}\gamma^\mu\hat{p}\hat{l}u
h_{\mu\nu}(\kappa)h_{\xi\sigma}(k_1)e^\rho
\Gamma_{\rho\nu\sigma}(k,\kappa,k_1)
\frac{1}{l_-}\frac{1}{2k_+(2p_+)^2}\cdot\frac{2(pl}{p_+}.\]
Calculating the matrix element and convoluting
$p^\xi p^\mu$ with the rest factors to form
the effective vertex $V$ (see (\ref{v}) we get
\[
M=2V\frac{1}{p_+^2k_+}.
\]
Since the time-factor contains only the term proportional to
$\kappa_+^2$ we need only the leading term from $V$:
The $T$ factor read from the diagram is
\[ T=\frac{8\kappa_+^2k_+}{\kappa_{\perp}^4k_\perp^2}.\]
The colour factor is
\[C=t^ct^dt^bf^{dca}=-\frac{1}{2}iNt^at^b\]
and thus
\[
(B4)=16iN(pl)t^at^b\frac{(k_1e)_{\perp}}{k_\perp^2}\frac{1}
{\kappa_{\perp}^2}.
\]
Integration over the angles of $\kappa$ transforms this
into
\beq
(B4)=16iN(pl)t^at^b\frac{(ke)_{\perp}}{k_\perp^2}\frac{1}{\kappa_{\perp}^2}.
\eeq

{\bf B41.}
The $M$-factor is the same with the substitution $V\to V_{1}$ (see Eq.
(\ref{v1}). The $T$-factor is
\[ T=\frac{8\kappa_+k_+^2}
{\kappa_\perp^2k_\perp^2(k_{1\perp}^2-k_\perp^2)}-
\frac{8\kappa_+^2k_+}{\kappa_\perp^4 k_\perp^2}.\]
Since it also contains  a term proportional
to $\kappa_+$ we have to take into account all  terms in $V_1$.
The colour factor is the same as in (B4)
So our final answer is
\beq
(B41)=16iN(pl)t^at^b\frac{(k_1e)_{\perp}}
{k_\perp^2\kappa_{\perp}^2}
\Big(1+\frac{(k_1\kappa)_\perp}{k_{1\perp}^2-k_{\perp}^2}\Big).
\eeq

{\bf B42.}
The $M$-factor is the same with the substitution $\kappa\to -\kappa$ in $V_1$
(\ref{v1}). Now the  $T$ factor contains only a term
proportional to $\kappa_+$ and we have to retain only the subleading
terms in the vertex.
The colour factor is the same as in (B4).
So in the end we find
\beq
(B42)=16iN(pl)t^at^b\frac{(k_1e)_{\perp}}
{k_\perp^2\kappa_{\perp}^2}
\frac{(k_1\kappa)}{k_{1\perp}^2-k_\perp^2}.
\eeq

{\bf B8.}
The momentum part of diagrams (B8), (B81) and (B82) is
the same as in (B4), (B41) and (B42) respectively.
In the colour factor $t^at^b\to t^bt^a$. So the difference
is only related with the $T$-factors.
For (B8)  the  $T$ factor
has the opposite sign as compared
to (B4). As a result we immediately get
\beq
(B8)=-16iN(pl)t^bt^a\frac{(ke)_{\perp}}{k_\perp^2}\frac{1}{\kappa_{\perp}^2}.
\eeq

{\bf B81.}
\beq
(B81)=-16iN(pl)t^bt^a\frac{(k_1e)_{\perp}}
{\kappa_\perp^2k_\perp^2k_{1\perp}^2}\Big(k_{1\perp}^2-k_\perp^2+
(k_1\kappa)_\perp\Big).
\eeq

{\bf B82.}
\beq
(B82)=-16iN(pl)t^bt^a\frac{(k_1e)_{\perp}}
{\kappa_\perp^2k_{1\perp}^2}\Big(1+\frac{(\kappa k_1)_\perp}{k_\perp^2}
\Big).
\eeq

{\bf B5}

Here we start with the colour factor.
\[C=f^{adc}t^ct^bt^d=
f^{adc}t^bt^ct^d+if^{adc}f^{cbe}t^et^d
=
f^{adc}t^ct^dt^b+if^{adc}f^{cbe}t^dt^e.\]
We have
\[ f^{adc}t^ct^d=-f^{acd}t^ct^d=-f^{adc}t^dt^c=
-i\frac{1}{2}Nt^a,\]
so that
\[C=\frac{1}{2}\Big [-i\frac{1}{2}N\{t^a,t^b\}+
if^{adc}f^{cbe}\{t^d,t^e\}\Big]\]
We use
\[ \{t^a,t^b\}=\frac{1}{N}\delta_{ab}+d^{abc}t^d
\]
to obtain
\[C=\frac{1}{2}\Big [-i\frac{1}{2}N\Big(
\frac{1}{N}\delta_{ab}+d^{abc}t^d\Big)+
if^{adc}f^{cbe}\Big(\frac{1}{N}\delta_{de}+d^{deh}t^h\Big)
\Big].\]
The identity
\[f^{adc}f^{cbe}d^{ehd}=-f^{dac}f^{cbe}d^{ehd}=
\frac{1}{2}Nd^{abh}\]
brings us to the final expression
\[C=\frac{1}{2}\Big [-i\frac{1}{2}
\delta_{ab}-i\frac{1}{2}Nd^{abd}t^d+
i\delta_{ab}+i\frac{1}{2}Nd^{abd}t^d\Big]=
i\frac{1}{2}\delta_{ab}.\]
So factor $C$ has a subdominant order in $N$ (the dominant
factor has order $N$).
This means that we can neglect all diagrams (B5), (B51)
and (B51).

{\bf B6.}
The $T$ factor has order $\kappa_+^3$, so that (B6)
can be neglected.

{\bf B61.}
We find the $M$-factor as
\[ M=-2i(pl)V_{1}\frac{1}{p_+^2k_+}.\]
The $T$ factor has order $\kappa_+^2$ so only the main
term from $V_1$ contributes.
The colour factor
\[ C=f^{adc_1}f^{c_1cb}t^ct^b.\]

{\bf B62.}
We have the same  $M$and $T$-factors.
The colour factor
\[C=f^{adc_1}f^{c_1bc}t^ct^b.\]
has the opposite sign as compared to (B61).

So $ (B6)+(B61)+(B62)=0$

{\bf B7.}
The $T$ factor has order $\kappa_+^3$ so that
$(B7)=0$

{\bf B71.}
As before
\[ M=-2i(pl)V_{1}\frac{1}{p_+^2k_+}.\]
The $T$ factor has order $\kappa_+$ so that only the subdominant
term in $V_1$ contributes
The colour factor is
\[ C=f^{ac_1c}f^{c_1db}t^ct^d. \]
It can be simplified:
\[ C=-[t^a,t^{c_1}][t^b,t^{c_1}]=
t^{c_1}t^at^bt^{c_1}+t^at^{c_1}t^{c_1}t^b-t^at^{c_1}t^bt^{c_1}-
t^{c_1}t^at^{c_1}t^b.\]
The last two terms are zero. The second term is equal to $(1/2)Nt^at^b$
The first can be calculated using
\[t^at^b=\frac{1}{2N}\delta_{ab}+F^{ab}_ct^c\]
with some numerical coefficients $F^{ab}_c$.
So the first term is $(1/4)\delta_{ab}$ and is negligible.
Thus
\[C=\frac{1}{2}Nt^at^b.\]

{\bf B72.}
Both the  $M$ and colour factors are the same.
So in the sum (B71)+(B72) the contribution  doubles

\beq
(B71)+(B72)=-32iN(pl)t^at^b\frac{(k_1\kappa)_{\perp}(k_2e)_{\perp}}
{\kappa_{\perp}^2k_{2\perp}^2(k_{1\perp}^2-k_{\perp}^2)}.
\eeq

{\bf B9.}
The $M$ factor is the same
\[M=-2i(pl)V\frac{1}{p_+^2k_+}.\]
The $T$ -factor is $\propto \kappa_+^2$ so only the first part
of $V$ contributes.
The colour factor is
\[ C=f^{ac_1b}f^{c_1dc}t^ct^d=-\frac{1}{2}iNf^{bac}t^c.\]

So
\beq
(B9)=-16(pl)Nf^{bac}t^c\frac{(k_1e)_{\perp}}
{\kappa_\perp^2k_{0\perp}^2}.
\eeq

{\bf B91.}
Again
\[ M=-2i(pl)V_{1}\frac{1}{p_+^2k_+}\]
Now  $T$ contains terms of the orders $\kappa_+$ and
$\kappa_+^2$. So both leading and subleading terms in $V_1$ contribute.
The colour factor is the same as in (B9). We find
\beq
(B91)=-16(pl)Nf^{bac}t^c\frac{(k_1e)_{\perp}}{\kappa_\perp^2k_{0\perp}^2}
\Big(1+\frac{(k_1\kappa)_\perp}{k_{1\perp}^2}\Big).
\eeq

{\bf B92.}
Again
\[ M=-2i(pl)V_{2}\frac{1}{p_+^2k_+}.\]
The $T$ factor contains only the term of the order $\kappa_+$.
So the subleading term in $V_2$ contributes.
The colour factor is the same as in (B9)
and we find
\beq
(B92)=-16(pl)Nf^{bac}t^c\frac{(k_1\kappa)_{\perp}(k_1e)_{\perp}}
{\kappa_{\perp}^2k_{0\perp}^2k_{1\perp}^2}.
\eeq

This ends the calculation of vertex part diagrams.

\subsection{Summary of the vertex diagrams}
We have obtained
the following non-zero contributions from the vertex diagrams.
Omitting the trivial common factor $-16iN(pl)$
\[
(B4)= -t^at^b\frac{(k_1e)_\perp}{k_\perp^2\kappa_\perp^2},
\]
\[
(B41)= -t^at^b\frac{(k_1e)_\perp}{k_\perp^2\kappa_\perp^2}
\Big(1+\frac{(k_1\kappa)_\perp}{k_{1\perp}^2-k_\perp^2}\Big),
\]
\[
(B42)= -t^at^b\frac{(k_1e)_\perp}{k_\perp^2\kappa_\perp^2}
\frac{(k_1\kappa)_\perp}{k_{1\perp}^2-k_\perp^2},
\]
\[
(B8)= t^bt^a\frac{(k_1e)_\perp}{k_\perp^2\kappa_\perp^2},
\]
\[
(B81)= t^bt^a\frac{(k_1e)_\perp}{k_\perp^2k_{1\perp}^2\kappa_\perp^2}
\Big(k_{1\perp}^2-k_\perp^2+(k_1\kappa)_\perp\Big),
\]
\[
(B82)= t^bt^a\frac{(k_1e)_\perp}{k_{1\perp}^2\kappa_\perp^2}
\Big(1+\frac{(k_1\kappa)_\perp}{k_\perp^2}\Big),
\]
\[
(B71)= t^at^b\frac{(k_2e)_\perp(k_1\kappa)_\perp}
{k_{2\perp}^2\kappa_\perp^2(k_{1\perp}^2-k_\perp^2)}\Big),
\]
\[
(B72)= t^at^b\frac{(k_2e)_\perp(k_1\kappa)_\perp}
{k_{2\perp}^2\kappa_\perp^2(k_{1\perp}^2-k_\perp^2)}\Big),
\]
\[
(B9)=-if^{bac}t^c\frac{(k_1e)_{\perp}}{\kappa_\perp^2k_{0\perp}^2},
\]
\[
(B91)=-if^{bac}t^c\frac{(k_1e)_{\perp}}{\kappa_\perp^2k_{0\perp}^2}
\Big(1+\frac{(k_1\kappa)_\perp}{k_{1\perp}^2}\Big),
\]
\beq
(B92)=-if^{bac}t^c\frac{(k_1e)_{\perp}}{\kappa_\perp^2k_{0\perp}^2}
\frac{(k_1\kappa)_\perp}{k_{1\perp}^2}.
\eeq

All terms involving the product
$(k_1\kappa)_\perp$ cancel between contributions of the diagrams
(*1) and (*2). This is nearly obvious, since in these diagrams
the direction of $\kappa_\perp$ is opposite. Indeed
compare (B41) and (B42)as an example. In (B41) $k_1+\kappa=k$, so
$(k_1\kappa)_\perp=(1/2)(k_\perp^2-k_{1\perp}^2-\kappa_\perp^2)$
The momentum part of the term with $(k_1\kappa)_\perp$ is
\beq
\frac{1}{2}\frac{(k_1e)_\perp}{k_\perp^2\kappa_\perp^2}
\Big(-1-\frac{\kappa_\perp^2}{k_{1\perp}^2-k_\perp^2}\Big).
\label{moma}
\eeq
The second term goes after the anglular integration. In terms of the integration
variable $k_{1\perp}$ we have $\kappa_\perp^2=(k-k_1)_\perp^2$.
In (B42) we have $k_1-\kappa=k$ so that
$(k_1\kappa)_\perp=-(1/2)(k_\perp^2-k_{1\perp}^2-\kappa_\perp^2)$
with $\kappa_\perp^2=(k_1-k)_\perp^2$.
So the momentum part of the term with $(k_1\kappa)_\perp$ will be the same
(\ref{moma}) with the opposite sign.
Thus the terms with $(k_1\kappa)_\perp$ cancel between (A41) and (A42).
The same is true for all vertex contributions.

After cancellation of these terms we find
\beq
(B4)+(B41)+(B42)= 2(B4)=
 32iN(pl)t^at^b\frac{(k_1e)_\perp}{k_\perp^2\kappa_\perp^2},
\label{sumb4}
\eeq
where we have added the suppressed overall factor,
\beq
(B8)+(B81)+(B82)= 2(B8)=
 -32iN(pl)t^bt^a\frac{(k_1e)_\perp}{k_\perp^2\kappa_\perp^2},
\label{sumb8}
\eeq
\beq
(B71)+(B72)=0
\eeq
and
\beq
(B9)+(B91)+(B92)= 2(B9)=
 -32iN(pl)if^{abc}t^c\frac{(k_1e)_\perp}{k_{0\perp}^2\kappa_\perp^2}.
\eeq
Since $k_1=k_0\pm\kappa$ and $k_0=k+q$
integration over the angles will change the last expression into
\beq
(B9)+(B91)+(B92)= 2(B9)=
 -32iN(pl)if^{abc}t^c\frac{[(k+q)e]_\perp}{(k+q)_\perp^2\kappa_\perp^2}.
\label{sumb9}
\eeq

Summing  (\ref{sumb4}), (\ref{sumb8}) and (\ref{sumb9}) we obtain our final
result for the contribution of vertex diagrams $S_v$
\[
S_v=(B4)+(B41)+(B42)+(B8)+(B81)+(B82)+ (B9)+(B91)+(B92)\]\beq=
- 32N(pl)f^{abc}t^c\frac{1}{\kappa_\perp^2}
\Big(\frac{(ke)_\perp}{k_\perp^2}-\frac{[(k+q)e]_\perp}{(k+q)_\perp^2}
\Big).
\eeq
Remarkably  this contribution seems to reverse the sign of the
similar contribution from the mass-terms. However we shall see that in fact
all vertex contributions are cancelled by renormalization.

\section{ Renormalization}
The perturbation theory we use is the unrenormalized one, where we write out all the diagrams
as they appear, including the self-masses in external lines. The only renormalization done is
that of the quark mass supposed to be zero, so that its self mass is also supposed
to be zero on mass-shell. No subtraction is initially done from the vertex part.

As a result, due to  self-mass insertions, the Born term is to be multiplied by the product of
factors $(Z-1)^{1/2}$ for each external line. So the lowest order diagrams, studied in Sec. 3.1
are to be multiplied by $(Z_q-1)(Z_g-1)^{1/2}$ where $Z_{q(g)}$ are the wave function renormalization
constant for the quark (gluon). In the following we denote
\[Z_{q(g)}-1=\xi_{q(g)}.\]
So we get an additional contribution
\beq
[(1)+(2)+(3)](\xi_q+\frac{1}{2}\xi_g).
\label{ap41}
\eeq
where (1),(2) and (3) are contributions of the lowest order presented in Sec. 3.3. Note that the
contribution (\ref{ap41}) is of the third order and, limiting to this order, we can substitute
the unrenormalized QCD coupling constant $g_0$ by the renormalized one $g$.

However this is not the whole story. In our unrenormalized perturbation
theory the Born terms carry the unrenormalized coupling constant $g_0$. Leaving aside the constant
associated with the external instantaneous interaction $L_0$, we have just one $g_0$ for gluon emission.
Recalling the relation
\beq
g_0=gZ_1Z_q^{-1}Z^g{-1/2}=g(1+\xi_1-\xi_q-\frac{1}{2}\xi_q).
\eeq
where $Z_1=1+\xi_1$ is the vertex renormalization constant, we further have to add  a term
\beq
[(1)+(2)+(3)](\xi_1-\xi_q-\frac{1}{2}\xi_g).
\label{ap42}
\eeq
In the sum of (\ref{ap41}) and (\ref{ap42}) contributions from self-masses cancel so that in the end
the additional contribution due to renormalization is just the Born term multiplied by the
vertex renormalization constant $\xi_1$
\beq
\xi_1[(1)+(2)+(3)].
\label{ap43}
\eeq
This means that we have to calculate the vertex part renormalization constant.

The definition of the latter is standard. Let $\Lambda^a_\rho(p,k)$ be the nontrivial vertex part with
the initial quark of momentum $p$, emitted gluon with momentum $k$ and Lorentz and colour indexes
$\rho$ and $a$ respectively. The vertex renormalization constant $\xi_1$ is defined by the value of
$\Lambda$ on mass shell:
\beq
\Big(\Lambda^a_\rho(p,k)\Big)_0=\xi_1t^a\gamma_\rho.
\label{ap44}
\eeq
The mass-shell can be defined in different ways depending on what is considered to be the
renormalized coupling constant. In our case the most natural definition is to choose the vertex
mass shell by putting all the three external particles on their mass shell
\beq
p^2=k^2=(p-k)^2=0.
\label{ap45}
\eeq
In the light cone variables with $p_0=p_\perp=0$,  to satisfy (\ref{ap45}) we have
to require $pk=p_+k_-=0$, that is
\[k_-=-\frac{k_\perp^2}{2k_+}=0, \]
which obviously requires $k_-=k_\perp^2=0$. So on the mass shell all three momenta $p,k$ and $p-k$ have
only non-zero plus components $p_+, k_+$ and $p_+-k_+$. In accordance with our kinematics we assume $k_+<<p_+$.

The simplest possibility to calculate $\xi_1$ is to use our previous results for the calculation of
the emission amplitude for diagrams with vertex insertions. Take diagrams (B8), (B81) and (B82).
In the Feynman language they correspond to the contribution
\beq
(B8)+(B81)+(B82)=t^b\bar{u}\frac{l}{l_-}\frac{\hat{p}-\hat{k}}{(p-k)^2}\Lambda^a_\rho(p,k)ue^\rho(k)
\cdot\frac{2(pl)}{p_+},
\label{ap46}
\eeq
the last factor coming from the target. The vertex part $\Lambda$ enters with two momenta on mass shell:
$p^2=k^2=0$, but the third momentum is off mass shell: $(p-k)^2=-2p_+k_-=k_\perp^2(p_+/k_+)$.
Putting $k_\perp^2\to 0$ in the vertex part we get the corresponding Born term multiplied by $\xi_1$:
\beq
\xi_1t^bt^a\bar{u}\frac{l}{l_-}\frac{\hat{p}-\hat{k}}{(p-k)^2}\hat{e}u\cdot\frac{2(pl)}{p_+}=
\xi_1\cdot(2),
\label{ap47}
\eeq
where (2) is the contribution from the lowest order diagram (2).
Note that the denominator $(p-k)^2$ vanishes as $k_\perp^2\to 0$, so that this limit can conveniently  be taken
with (\ref{ap46}) multiplied by $(p-k)^2$
\beq
k_\perp^2\frac{p_+}{k_+}[(B8)+(B81)+(B82)]_{k_\perp\to 0}=
\xi_1k_\perp^2\frac{p_+}{k_+}\cdot(2).
\eeq
The l.h.s of this equality can be read off Eq.(103):
\beq
l.h.s=\frac{p_+}{k_+}32N(pl)t^bt^a\frac{(ke)_\perp}{\kappa_\perp^2},
\eeq
where as  always we suppressed integration over $\kappa_+$ and $\kappa_\perp$ with the usual weight.
The r.h.s can be read off Eq. (70);
\beq
r.h.s=\xi_1\cdot 8t^bt^a(pl)(ke)_\perp,
\eeq
wherefrom we conclude
\beq
\xi_1=4N\frac{1}{\kappa_\perp^2}.
\label{ap48}
\eeq

This result is confirmed by direct calculations using the Feynman diagram approach.

We see that in our approximation the vertex part does not depend on $k_\perp^2$ at all.
This means that renormalization will eliminate all vertex contributions to our emission amplitude.
As a result, the only contribution which remain come from the self-mass insertions.

\section{Reggeization of the interaction}

After cancellation of all contributions from the vertex part insertions we are left with
the contributions from the self-mass insertions, which we study here.
We first combine terms (A1)-(A4)
Summing (A1)+(A3) we find
\beq
(A1)+(A3)=16N(pl)f^{abc}t^c\frac{(ke)_{\perp}}
{k_\perp^2}\frac{1}{\kappa_\perp^2}.
\eeq
Summing (A2)+(A4)
\beq
(A2)+(A4)=-16N(pl)f^{abc}t^c\frac{(ke)_{\perp}}
{k_\perp^2}\frac{[\kappa(\kappa+q)]}{\kappa_\perp^2(\kappa+q)_\perp^2}.
\eeq
Thus
\beq
S_1\equiv\sum_{j=1}^4(A_j)=
16N(pl)f^{abc}t^c\frac{(ke)_{\perp}}
{k_\perp^2}\Big(\frac{1}{\kappa_\perp^2}-
\frac{[\kappa(\kappa+q)]}{\kappa_\perp^2(\kappa+q)_\perp^2}\Big).
\eeq
Under the sign of integration over $\kappa$ we can rewrite this as
\[
S_1=
8N(pl)f^{abc}t^c\frac{(ke)_{\perp}}
{k_\perp^2}
\Big(\frac{1}{\kappa_\perp^2}+\frac{1}{(\kappa+q)_\perp^2}-
2\frac{[\kappa(\kappa+q)]}{\kappa_\perp^2(\kappa+q)_\perp^2}\Big)\]
\beq
=8(pl)f^{abc}t^c\frac{(ke)_{\perp}}{k_\perp^2}
\frac{Nq^2}{\kappa_\perp^2(\kappa+q)_\perp^2}.
\eeq
Integration over $\kappa$ with the additional factor $g^2$ converts
the last factor into $y\omega(q)$ where $y$ is the harder gluon rapidity and
$\omega(q)$ is the gluon trajectory Regge (see Eq. (\ref{omy})). So we have found
\beq
S_1=8(pl)f^{abc}t^c\frac{(ke)_{\perp}}{k_\perp^2}\cdot y\omega(q)
\eeq

Now we turn to the rest three diagrams (A5), (A6) and (A7).
We find
\[ (A6)+(A7)=-(A6) \]
and as a result
\beq
S_2=\sum_{j=5}^7(A_j)=
16N(pl)f^{abc}t^c\frac{[(k+q)e]_{\perp}}{(k+q)_\perp^2}
\Big(\frac{[\kappa(\kappa+q)]_{\perp}}{\kappa_{\perp}^2(\kappa+q)_\perp^2}-
\frac{1}{\kappa_\perp^2}\Big).
\label{a567}
\eeq
Under the sign of integration over $\kappa$ this can be rewritten as
\[
S_2=
8N(pl)f^{abc}t^c\frac{[(k+q)e]_{\perp}}{(k+q)_\perp^2}
\Big(2\frac{[\kappa(\kappa+q)]_{\perp}}{\kappa_{\perp}^2(\kappa+q)_\perp^2}-
\frac{1}{\kappa_\perp^2}-\frac{1}{(\kappa+q)_\perp^2}\Big)\]\beq=
-8(pl)f^{abc}t^c\frac{[(k+q)e]_{\perp}}{(k+q)_\perp^2}
\frac{Nq^2}{\kappa_{\perp}^2(\kappa+q)_\perp^2}.
\label{a567a}
\eeq
Integration over $\kappa$ wit factor $g^2$ then gives
\beq
S_2=-8(pl)f^{abc}t^c\frac{[(k+q)e]_{\perp}}{(k+q)_\perp^2}.
\cdot y\omega(q)
\eeq

In conclusion the sum of all self-mass diagrams except
the last term in (\ref{a7}) gives the contribution
\beq
\sum_{j=1}^7(A_j)=8(pl)f^{abc}t^c\Big(\frac{(ke)_{\perp}}{k_\perp^2}
-\frac{[(k+q)e]_{\perp}}{(k+q)_\perp^2}\Big)\cdot y\omega(q).
\eeq
This is just the expected expression implying reggeization of the
interaction with the target.

\section{Conclusions}
We have studied corrections to the lowest order emission amplitudes for the quark
moving in the external instantaneous potential and emitting a soft gluon. Only
corrections due to virtual gluons softer than the emitted one have been taken into account.
The reason is that they play a decisive role in the study of the evolution of the
inclusive cross-section for gluon production at rapidities smaller than that of the observed gluon,
where the discrepancy between the results of ~\cite{kovch} and ~\cite{BSV} originates.
We have found that these corrections are fully equivalent to the corresponding corrections in
the BFKL language, where they come from the gluon reggeization
(and in fact from very different Feynman graphs). Renormalization has played a decisive role
in obtaining this result.

We have to stress that in principle this result is insufficient to establish full equivalence
of the physical picture used in  ~\cite{kovch} and ~\cite{BSV}. The difference
between their results starts from the next-to-leading order, where a huge number of more
complicated diagrams  appear with a greater number of interactions (two or three). The study of loops in such
diagrams, although straightforward, requires an extraordinary effort, for which we are not ready at
present. However results obtained in this work lead us to believe that also in this
more complicated case the results of both approaches coincide.

As mentioned in the introduction, the  equivalence of loop diagrams, if fully established,
together with
the already proved equivalence of tree diagrams shows that the physical foundations
on which derivations in ~\cite{kovch} and ~\cite{BSV} are based are the same. The difference
in their results can therefore either be in fact absent, if the new terms  in ~\cite{BSV}
cancel, or originate from the difference in the derivations
themselves. The study of the latter difference is not simple, since the technique used is
completely different. The final answer therefore requires a detailed analysis of the procedures
used, which  is postponed for future studies. The immediate and simpler task is to check if
the terms with the BKP states found in ~\cite{BSV} are actually present and not cancelled.
Our preliminary studies indicate that they are not. After due verification this result
will be  presented for a future publication.

\section{Acknowledgments}
The author is deeply indebted to Yu.Kovchegov for numerous illuminating and
constructive discussions. The author is most thankful to B.Vlahovic for his
interest for this work and support during the author's stay at NCCU in Durham, NC, USA,
where part of this study was done. This work has been also supported by grants
RFFI 09-012-01327-a and RFFI-CERN of Russia.

\section{Appendix 1. Calculation of the vertex insertion}

Consider  the diagram Fig. \ref{fig9}. At the first step we convolute the 3-gluon
 vertex with the polarization vector $E$.
We take into account that
$
\kappa_{1\nu}H^{\mu\nu}(\kappa_2\kappa_1)=0,
$
so that in the vertex we may drop terms proportional to $\kappa_{1\nu}$
and $k_{1\sigma}$ to find
\beq
\Gamma_{\nu\sigma}\equiv
\Gamma_{\nu\sigma\rho}(\kappa_1,k_1,k_0)E^{\rho}(k_0)
=2k_{1\nu}E_{\sigma}-2\kappa_{1\sigma}E_{\nu}+g_{\nu\sigma}(\kappa_1-k_1,E).
\label{aeq}
\eeq

Next we convolute $H_{\mu\nu}(\kappa_2\kappa_1)$ with $p^\mu k_1^\nu$
\beq
A_1\equiv p_\mu k_{1\nu}H^{\mu\nu}(\kappa_2,\kappa_1)=
\frac{p_+k_{+}}{\kappa_+^2}(\kappa_1\kappa_2)_{\perp}-
\frac{p_+}{\kappa_+}(k_1\kappa_2)_\perp
\eeq
and $p_\mu E_\nu$
\beq
A_2\equiv p_\mu E_\nu H^{\mu\nu}(\kappa_2,\kappa_1)=
-\frac{p_+}{\kappa_+}(\kappa_2e)_{\perp}.
\eeq

Interchanging $\kappa_{1,2}\lra k_{1,2}$ we also find
\beq
A_3\equiv \kappa_{1\sigma}H^{\sigma}(k_2,k_1)
=\frac{p_+\kappa_{+}}{k_+^2}(k_1k_2)_{\perp}-
\frac{p_+}{k_+}(\kappa_1k_2)_\perp
\eeq
and
\beq
A_4
\equiv E_\sigma
H^{\sigma}(k_2,k_1)=
-\frac{p_+}{k_+}(k_2e)_{\perp}.
\eeq

We have
\beq
V=2A_1A_4-2A_2A_3+\tilde{V},
\label{vtot}
\eeq
where
\beq
\tilde{V}=(\kappa_1-k_1,E)H_{\nu}(\kappa_2,\kappa_1)H^\nu(k_2,k_1).
\eeq
After some calculations we find
\beq
\tilde{V}=2\frac{p_+^2}{\kappa^++k_+}(\kappa_2k_2)_{\perp}
\Big(\frac{(\kappa_1e)_{\perp}}{\kappa_+}-\frac{k_1e)_{\perp}}{k_{1+}}\Big).
\eeq

Summing terms in (\ref{vtot}) we take into account that $\kappa_+<<k_+$
and retain only the leading and the next-to-leading terms to find
(\ref{v}).

\section{Appendix 2. Calculation of the gluon-self mass insertion}
We start calculating
$
A^{(1)}_{\mu\sigma}\equiv H_{\mu\nu}\Gamma_{\nu\sigma},
$
where $\Gamma_{\nu\sigma}$ was defined in (\ref{aeq}).
We define
\beq
b_{1\mu}\equiv H_{\mu\nu}(\kappa_2,\kappa_1)k_1^\nu=
k_{1\mu}-\frac{l_{\mu}(k_1\kappa_2)+\kappa_{1\mu}(k_1l)}{(\kappa l)}+
l_\mu\frac{(k_1l)(\kappa_2\kappa_1)}{(\kappa l)^2}
\eeq
and
\beq
b_{2\mu}\equiv H_{\mu\nu}(\kappa_2,\kappa_1)E^\nu =
E_{\mu}-l_\mu\frac{(\kappa_2E)}{(\kappa l)},
\eeq
which gives
\[
A^{(1)}_{\mu\sigma}=2E_\sigma b_{1\mu}-2\kappa_{1\sigma}b_{2\mu}
+[(\kappa_1-k_1)E]H_{\mu\sigma}(\kappa_2\kappa_1).
\]

Next we multiply this by $H_{\xi\sigma}(k_2,k_1)$:
\beq
A^{(2)}_{\xi\mu}\equiv H_{\xi\sigma}(k_2,k_1)A^{(1)}_{\mu\sigma}.
\eeq
We  define
\beq
a_{1\xi}\equiv H_{\xi\sigma}(k_2,k_1)E^\sigma=E_{\xi}-
l_\xi\frac{k_2E}{(kl)},
\eeq
\beq
a_{2\xi}\equiv H_{\xi\sigma}(k_2,k_1)\kappa_1^\sigma=
\kappa_{1\xi}-\frac{l_{\xi}(k_2\kappa_1)+k_{1\xi}(\kappa_1l)}{(kl)}
+l_{\xi}\frac{(\kappa_1l)(k_2k_1)}{(kl)^2}
\eeq
and
\beq
\tilde{H}_{\xi\mu}(k_2,\kappa_2)\equiv
H_{\xi\sigma}(k_2,k_1)H_{\mu\sigma}(\kappa_2,\kappa_1)
=g_{\xi\mu}-\frac{l_\xi k_{2\mu}}{(kl)}-
\frac{l_\mu\kappa_{2\xi}}{(\kappa l)}+\frac{l_\mu l_\xi (\kappa_2k_2)}
{(\kappa l)(kl)}.
\eeq
This gives
\beq
A^{(2)}_{\xi\mu}=2a_{1\xi}b_{1\mu}-2a_{2\xi}b_{2\mu},
+ \tilde{A}^{(2)}_{\xi\mu}
\eeq
where
\beq
\tilde{A}^{(2)}_{\xi\mu}=[(\kappa_1-k_1)E]\tilde{H}_{\xi\mu}(k_2,\kappa_2).
\eeq

At the next step we calculate
\beq
A^{(3)}_{\xi\mu}\equiv  H^{\beta}(k_4,k_3)
\Gamma_{\beta\xi\mu}(k_3,k_2,\kappa_2),
\eeq
where
\beq
H^\beta (k_4,k_3)=p_\alpha H^{\alpha\beta}(k_4,k_3).
\eeq
We find
\beq
A^{(3)}_{\xi\mu}=2\kappa_{2\xi}H_{\mu}(k_4,k_3)-2k_{2\mu}H_\xi(k_4,k_3)
+\tilde{A}^{(3)}_{\xi\mu},
\eeq
where
\beq
\tilde{A}^{(3)}_{\xi\mu}=g_{\mu\xi}(k_2-\kappa_2)_{\beta}H^{\beta}(k_4,k_3)
=
g_{\xi\mu}
\Big(2\frac{p_+}{k_+}(\kappa_2k_4)_{\perp}-
\frac{p_+\kappa_+}{k_+^2}(k_3k_4)_{\perp}\Big).
\label{ta3}
\eeq

Multiplying $A^{(2)}$ by $A^{(3)}$ we obtain $\Pi$ as a sum of 9 terms:
\[
\Pi=4(a_1\kappa_2)(b_1H)-4(a_2\kappa_2)(b_2H)+2\kappa_{2\xi}
\tilde{A}^{(2)}_{\xi\mu}H_\mu\]\[
-4(a_1H)(k_2b_1)+4(a_2H)(k_2b_2)-2H_{\xi}\tilde{A}^{(2)}_{\xi\mu}k_{2\mu}
+2a_{1\xi}\tilde{A}^{(3)}_{\xi\mu}b_{1\mu}\]\beq-
2a_{2\xi}\tilde{A}^{(3)}_{\xi\mu}b_{2\mu}
+\tilde{A}^{(2)}_{\xi\mu}\tilde{A}^{(3)}_{\xi\mu}.
\label{pi1}
\eeq
Since $(a_1l)=(a_2l)=(b_1l)=(b_2l)=0$ we can drop terms proportional to
$l$ in $H$, so that for our purpose
\beq
H_{\beta}=p_{\beta}-k_{4\beta}\frac{(pl)}{(kl)}.
\eeq

Next we study the 9 terms in (\ref{pi1}) successively, denoting them as
(1), (2) and so on. We find in  orders in $1/\kappa_+^2$ and $1/\kappa_+$
\[
(1)=4(a_1\kappa_2)(b_1H)=4\frac{p_+}{\kappa_+}(\kappa_1k_4)_{\perp}(\kappa_2e)_{\perp},
\]
\[
(2)=-4(a_2\kappa_2)(b_2H)=0,
\]
\[
(3)=
2\kappa_{2\xi}\tilde{A}^{(2)}_{\xi\mu}H_{\mu}=0,
\]
\[
(4)=-4(a_1H)(b_1k_2)=
4\frac{p_+k_+}{\kappa_+^2}(\kappa_1\kappa_2)_{\perp}(k_4e)_{\perp}-
4\frac{p_+}{\kappa_+}(k_4e)_{\perp}
\Big((k_1\kappa_2)_{\perp}+(k_2\kappa_1)_{\perp}\Big),
\]
\[
(5)=4(a_2H)(b_2k_2)=
4\frac{p_+}{\kappa_+}(k_4\kappa_1)_{\perp}(\kappa_2e)_{\perp},
\]
\[
(6)=-2H_{\xi}\tilde{A}^{(2)}_{\xi\mu}k_{2\mu}=
-4\frac{p_+}{\kappa_+}(\kappa_2k_4)_{\perp}(\kappa_1e)_{\perp},
\]
\[
(7)=2a_{1\xi}\tilde{A}^{(3)}_{\xi\mu}b_{1\mu}=(6),
\]
\[
(8)=-2a_{2\xi}\tilde{A}^{(3)}_{\xi\mu}b_{2\mu}=0,
\]
\beq
(9)=\tilde{A}^{(2)}_{\xi\mu}\tilde{A}^{(3)}_{\xi\mu}=0
\eeq
Terms which are written as equal to zero in fact are finite at $\kappa_+=0$
and so can be neglected.

Summing all terms we finally obtain (\ref{pi}).

\end{document}